%% file: main_draft.tex
\PassOptionsToPackage{table}{xcolor}

\documentclass[11pt]{article}

\usepackage[preprint]{acl}

\usepackage[T1]{fontenc}
\usepackage[utf8]{inputenc}

\usepackage{times}
\usepackage{latexsym}
\usepackage{microtype}
\usepackage{inconsolata}

\usepackage[table]{xcolor}      % for \rowcolor
\usepackage{array}
\usepackage{amssymb}            % for \blacktriangle, \blacktriangledown
\usepackage{booktabs}
\usepackage{caption}
\usepackage{cuted}
\usepackage{float}
\usepackage{multirow}
\usepackage{graphicx}
\usepackage{enumitem}
\usepackage{amsmath,amssymb}
\usepackage{hyperref}
\usepackage{siunitx}
\usepackage{subcaption}
\usepackage[most]{tcolorbox}
\usepackage{todonotes}

\input{macros}
 \title{On the impact of retrieved content representations in RAG pipelines\thanks{Code available at \url{https://github.com/ielab/rag-representation}}}

\author{
	Jonathan J Ross \\
	The University of Queensland \\
	\texttt{jonathan.ross@uq.edu.au} \\\And
	Bevan Koopman \\
	CSIRO / The University of Queensland \\
	\texttt{bevan.koopman@csiro.au} \\\AND
	Anton van der Vegt \\
	The University of Queensland \\
	\texttt{a.vandervegt@uq.edu.au} \\\And
	Guido Zuccon \\
	The University of Queensland \\
	\texttt{g.zuccon@uq.edu.au} \\
}
\date{}  % leave empty

\begin{document}
	\nolinenumbers
	\maketitle
	
	% ---------- Content ----------
	\input{content}

	% ---------- Bibliography ----------
	\bibliography{bibliography/references}
	
	% ----- Appendix ---------
	\appendix
	\onecolumn
	\input {appendix}

\end{document}

%% file: content.tex
% content.tex
% Defines the structure and order of the paper.
% Contains NO prose.

\input{sections/01_abstract}
\input{sections/02_introduction}

\input{sections/03_related_work}

\input{sections/04_experiments}

\input{sections/05_results}

\input{sections/06_discussion}

\input{sections/07_conclusion}

\input{sections/08_limitations}

%% file: sections/01_abstract.tex
\begin{abstract}
Retrieval-Augmented Generation (RAG) supplements a language model's input with retrieved documents, yet most RAG pipelines inherit retrieval components designed for human readers. How retrieved content should be represented when the consumer is a large language model (LLM) rather than a human is less well understood. Recent work has proposed transformations of retrieved content and identified properties that affect generation, but each examines a single transformation or property in isolation, leaving open which features of a document's representation matter most. We address this with a controlled comparison: holding retrieval fixed, we vary only the representation of retrieved documents, comparing an original baseline against thirteen transformations spanning selection, summarisation, and reformulation, in query-dependent and query-independent variants. Across these fourteen representations we measure question-answering accuracy for four generators, and for each representation we also measure answer retention: whether a known answer-bearing document still supports its answer after transformation. We find that answer retention is the primary determinant of generator accuracy; notably, when retention is high, a representation's wording, structure, length, and query-dependence have limited effect. This suggests that accuracy gains attributed to specific mechanisms in prior work may be partly explained by how well those mechanisms preserve answer-bearing content, an attribution that cannot be settled without controlling for retention.
\end{abstract}

%% file: sections/02_introduction.tex
\section{Introduction}

Retrieval-Augmented Generation (RAG) supplements the input of a large language model (LLM) with external documents sourced by a retrieval system \cite{lewis_retrieval-augmented_2020, guu_realm_2020}. RAG has enabled strong performance on knowledge-intensive tasks \cite{izacard_atlas_2023, wei_browsecomp_2025}, improved answer attribution \cite{bohnet_attributed_2023}, and allowed models to incorporate new knowledge without retraining. 

Most RAG systems adopt retrieval components built to serve human users, and many of their design choices reflect that origin: ranked result lists \cite{joachims_clickthrough_2005}, query-biased snippets \cite{tombros_advantages_1998}, and query-term highlighting \cite{iofciu_evaluating_2009} all help a human reader locate and consume relevant information. These design choices are grounded in decades of study of how human readers process search results, and RAG pipelines have largely inherited them.

What works best when the consumer is an LLM is less well understood, but there are consequential differences.  For example, LLMs are sensitive to changes in input text that would not be meaningful to a human reader: \citet{sclar_quantifying_2024} found that prompts differing in minor formatting choices such as separators or capitalisation produced accuracy differences of up to 76 points. Unrelated input affects answer generation in unintuitive ways: \citet{cuconasu_power_2024} found that randomly sampled passages can actually improve RAG accuracy. And beyond accuracy, the cost of processing retrieved content scales superlinearly with its length, giving practical weight to representation choices that a human consumer would not require. 

If LLMs process retrieved content differently from humans, the representation in which that content is presented to them --- how it is worded and structured, and not just what information it contains --- becomes a design choice worth examining directly. Recent work has begun to study how the representation of retrieved content affects downstream generation.  Recursive clustering and summarisation of passages produces a tree spanning multiple levels of abstraction \citep{sarthi_raptor_2024}; post-retrieval compression reduces input length while preserving accuracy \citep{xu_recomp_2024}; and document rewriting optimises retrieved content for generative utility rather than retrieval relevance \citep{kim_relevance_2025}. A separate line of work has identified specific properties of retrieved content that affect generator behaviour, including query similarity and semantic completeness \citep{tan_blinded_2024}, and relevance combined with logical interconnectivity \citep{chang_what_2025}. These studies demonstrate that representation matters for downstream performance, but each examines a single technique or property in isolation. What is missing is a controlled comparison: varying the representation of retrieved content, holding other factors fixed, to identify which properties matter most when the consumer is an LLM rather than a human.

We address this gap by treating the representation as a design choice and studying it directly. Holding retrieval fixed, we vary only the representation of retrieved content via a set of document representation transformations spanning selection, summarisation, and reformulation, with both query-dependent and query-independent variants. We then examine the effect on answer accuracy. For each representation, we also measure answer retention: whether a known answer-bearing document still contains a valid answer after transformation.

We find that answer retention is the primary determinant of generator accuracy. Other aspects of the representation --- its wording, structure, length, and whether it incorporates the query --- have limited effect when retention is high, suggesting that recent gains attributed to specific representation mechanisms may be partly explained by how well those mechanisms preserve answer-bearing content. Our contributions are as follows:
\begin{itemize}
	\item We conduct a controlled comparison of fourteen document representations---an \emph{original} baseline and thirteen transformations spanning selection, summarisation, and reformulation---holding retrieval fixed to isolate the effect of representation on generator accuracy.
	\item We find that answer retention is the primary determinant of generator accuracy across representations.
	\item We show that query-dependent representations do not systematically outperform query-independent ones.
	\item We find that generators do not prefer LLM-produced transformations over non-LLM ones; apparent preferences track answer retention rather than transformation source.
	\item We argue that accuracy gains cannot be attributed to specific representation mechanisms without controlling for answer retention, which prior work has not generally done.
\end{itemize}

Our findings provide empirical grounding for the design of retrieval systems that serve LLMs rather than humans.

%% file: sections/03_related_work.tex
\section{Related Work}

A growing body of work asks how the representation of retrieved content affects generator accuracy in RAG. Two directions have emerged. The first, transformation-based, proposes specific transformations of retrieved content and measures their effect on accuracy. The second, property-based,  characterises which properties of retrieved content generators are sensitive to.

Transformation-based studies intervene at one of two points in the pipeline: at indexing, by changing what gets stored and retrieved; or post-retrieval, by transforming retrieved content before it reaches the generator.   At indexing, \citet{chen_dense_2024} use atomic propositions rather than passages, arguing that propositions are a more effective unit of retrieval granularity, and RAPTOR \cite{sarthi_raptor_2024} builds hierarchical summaries to make different levels of abstraction retrievable. Post-retrieval, HtmlRAG \cite{tan_htmlrag_2025} preserves HTML structure rather than flattening retrieved content to plain text, arguing that structural information aids the generator. Compression methods \cite{xu_recomp_2024, pan_llmlingua-2_2024, li_compressing_2023} reduce retrieved content to lower inference cost, finding that generators tolerate substantial reductions with comparable or only slightly degraded accuracy.  Finally, rewriting methods learn to transform retrieved content for downstream generator utility rather than for retrieval relevance \cite{kim_relevance_2025, li_align_2026}. Each of these reports an accuracy effect --- a gain, or preservation under compression --- and attributes it to the proposed mechanism, but does not isolate which properties of the resulting representation drive that effect.

Several properties of retrieved content shape generator behaviour. Utility --- whether a passage actually improves a generator's answer --- is a property distinct from retriever-side relevance \cite{tian_is_2025}, and one that varies by generator: \citet{zhang_llm-specific_2025} show that passages maximising utility for one generator are consistently suboptimal for another. Logical connectivity across retrieved passages also matters: \citet{chang_what_2025} find that LLMs prefer external knowledge that is both relevant and logically connected. Irrelevance also matters: \citet{cuconasu_power_2024} find that distracting irrelevance degrades accuracy where random irrelevance does not, and \citet{amiraz_distracting_2025} formalise this as a continuous, per-passage measure. Most directly, sufficient context --- whether retrieved content contains enough information for a plausible answer --- predicts generator accuracy without fully determining it \cite{joren_sufficient_2025}.  Each of these properties has been studied in isolation, without comparison against others.

An additional property of retrieved content is its source: whether it was produced by a human or an LLM, and if by an LLM, which LLM. Source sits one step removed from the properties above: it is not detected by the generator directly but imbues retrieved content with detectable properties, such as stylistic markers, that can influence generator behaviour.  Outside RAG, LLMs favour LLM-generated text when choosing between human- and LLM-authored options \citep{laurito_aiai_2025}, and favour self-generated content over other-model content in evaluation \citep{panickssery_llm_2024}. Within RAG, the picture is mixed. \citet{tan_blinded_2024} show that when generated and retrieved contexts conflict, LLMs systematically favour the generated one, even when it is incorrect. In contrast, \citet{chen_llms_2025} find that in fact-centric RAG, this self-preference disappears, and factual accuracy --- not whether the content is human- or LLM-authored --- shapes the generator's output. Whether this preference for LLM-produced content extends to the representation of retrieved documents in a full RAG pipeline remains open.

Together, these lines show that the representation of retrieved content matters for generation and that distinct properties of that representation drive generator behaviour, but each has been examined in isolation. A controlled comparison that places these properties in competition is the approach we take.

%% file: sections/04_experiments.tex
% experimental methooology]
\begin{figure}[t]
	\centering
	\includegraphics[width=\columnwidth]{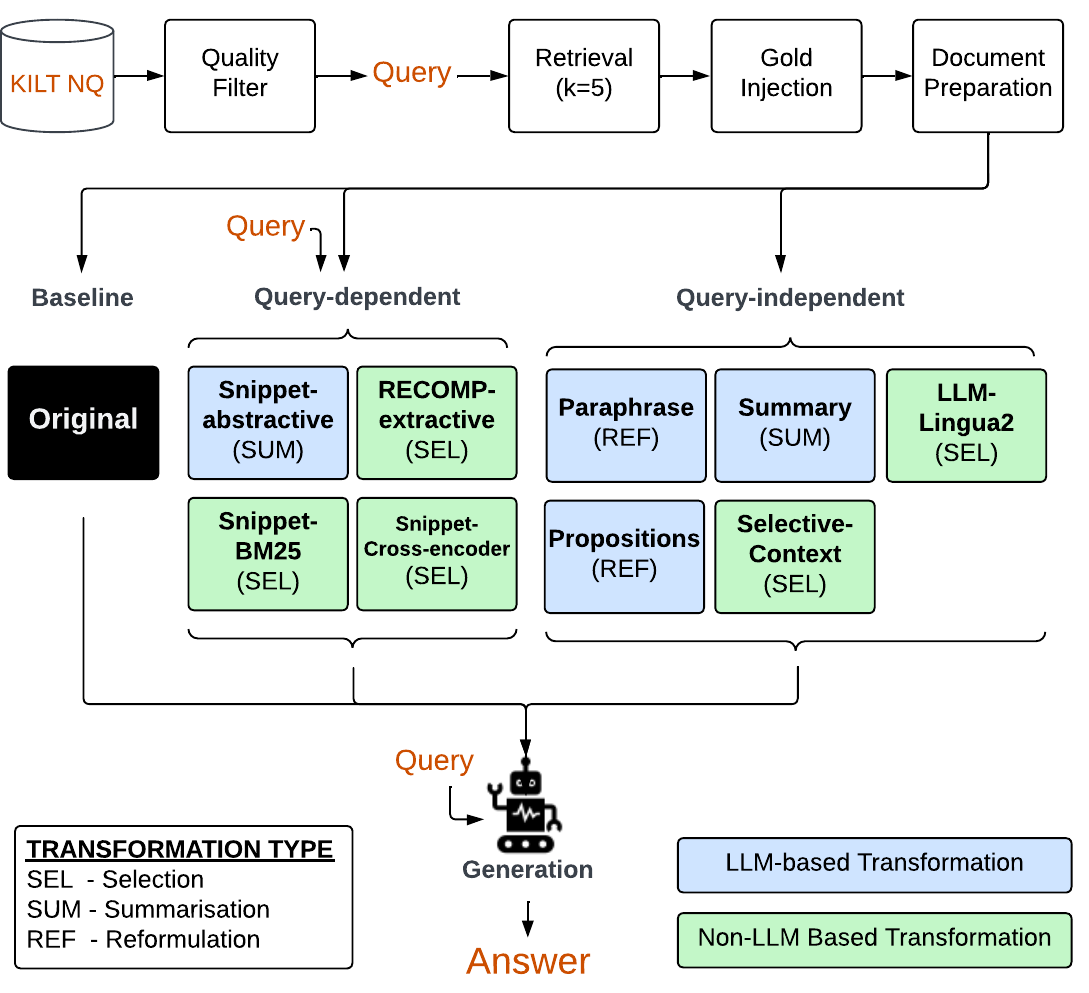}
	\caption{Overview of the experimental pipeline. KILT-NQ queries are filtered and used to retrieve five documents; gold injection ensures an answer-bearing document is present in every retrieved set, holding retrieval fixed so that downstream differences are attributable to representation rather than retrieval. Each retrieved document is passed to the generator either unchanged (Original) or after one transformation, organised by transformation type (selection, summarisation, reformulation; SEL/SUM/REF) and by whether the transformation incorporates the query (query-dependent vs. query-independent). Blue boxes denote LLM-based transformations, green boxes non-LLM ones. The ten boxes comprise the Original baseline and nine transformation methods; the four LLM-based methods are each run with two transformation models (Gemma 3 27B, Llama 3.3 70B), giving the fourteen representations evaluated (Table~\ref{tab:rep_properties})}
	\label{fig:method_overview}
\end{figure}

\section{Experimental Setup}
\label{sec:experimental_setup}

To isolate the effect of document representation on RAG question answering, we fix the retrieval and vary only the representation of retrieved documents, which are either passed directly to the generator or first transformed into an alternative representation (Figure~\ref{fig:method_overview}).  For each query, five documents are retrieved, and the generator produces a short-form answer. We ensure that at least one retrieved document --- the gold document, which contains an answer for that query --- is present in every retrieved set, so that differences in answer quality are attributable to representation, not retrieval of relevant content.

Three reference conditions provide additional points of comparison. In the \emph{closed-book} condition, the generator answers without any retrieved documents. Two oracle conditions provide further reference points: \emph{gold-only} passes only the gold document to the generator, isolating performance when the answer-bearing document is presented without additional retrieved content; \emph{gold-5x} passes five copies of the gold document, increasing input length and adding repetition while holding the information content fixed. Comparing gold-5x to gold-only isolates the effect of length and repetition; comparing it to the original retrieved set isolates the effect of non-gold retrieved content.

\subsection{Dataset and Retrieval}
\label{subsec:dataset}

 Our dataset selection was guided by three requirements. First, we required a single-hop, short-answer question answering task so that changes in answer quality can be attributed to document representation, not multi-hop reasoning. Second, we required source texts with section-level structure, since several transformations (e.g., section summaries, per-section proposition extraction) operate at that granularity. Third, we required gold document annotations so that a known answer-bearing document could be substituted into the retrieved set, ensuring every query had an answerable context.

KILT-NQ met all three criteria \cite{petroni_kilt_2021}. It pairs Natural Questions queries with a frozen Wikipedia corpus and annotates each query with a gold answer set (the acceptable answer strings) and gold page annotations (the Wikipedia pages containing those answers), so that annotations and retrieval operate over the same document collection. It also provides per-answer provenance, mapping each answer in the gold answer set to its specific source page; we refer to such a source page as a gold document. This is necessary because a single query may have valid answers on different gold documents; without per-answer provenance, a gold document already in the retrieved set could go unrecognized, resulting in unnecessary substitution and an inconsistent number of gold documents across queries.

Queries were selected from the KILT-NQ validation set. From the original 2,837 queries, 446 were removed in two filtering steps. First, 87 queries were excluded for lacking a gold answer or gold-document annotation entirely.  Second, 359 queries were removed because their gold annotations, though present, could not support reliable evaluation, for one of two reasons. Some had no gold document containing an answer from the gold answer set, leaving answer retention undefined and breaking the design's reliance on an answer-bearing gold document. Others had correct answers absent from the gold answer set, which would cause the evaluator to mark correct generations as incorrect. The remaining 2,391 queries were used for this study.

Retrieval was performed against the KILT Wikipedia corpus. The $\sim$92.3M paragraphs from the corpus were encoded with BAAI/bge-small-en-v1.5 and retrieved against queries using cosine similarity (FAISS flat, exact search).  Because transformations operate at the page level, retrieved paragraphs were mapped to their source Wikipedia pages, and the first five unique pages encountered in rank order were selected as the retrieved documents.  Multiple paragraphs often map to the same page, so for each query the top 100 paragraphs were retrieved so that at least five unique pages were available.  To ensure that every query had a gold page in the retrieved set, the retrieved documents were checked against the per-answer provenance annotations. Where no gold page was present, one was substituted for the fifth-ranked document. Of the 2,391 queries, 933 (39\%) required substitution.

% document representation properties
\begin{table}[t]
	\centering
	\setlength{\tabcolsep}{4pt}
	{\footnotesize
		\begin{tabular}{@{}>{\raggedright\arraybackslash}p{3.7cm}>{\raggedright\arraybackslash}p{4.0cm}@{}}
			\toprule
			\textbf{Representation} & \textbf{Method}  \\
			\midrule
			Original (baseline)             & None    \\
			\midrule
			\multicolumn{2}{@{}l}{\textit{Selection}} \\
			\hspace{0.5em} Snippet-bm25              & BM25 sentence scoring      \\
			\hspace{0.5em} Snippet-cross-encoder     & Cross-encoder sentence scoring     \\
			\hspace{0.5em} RECOMP-extractive-50      & Bi-encoder sentence scoring         \\
			\hspace{0.5em} LLMLingua2-50             & Token classification \& pruning  \\
			\hspace{0.5em} Selective-context-50      & Surprisal-based sentence pruning        \\
			\midrule
			\multicolumn{2}{@{}l}{\textit{Summarisation}} \\
			\hspace{0.5em} Summary-Gemma             & Section summarisation      \\
			\hspace{0.5em} Summary-Llama             & Section summarisation        \\
			\hspace{0.5em} Snippet-abstractive-Gemma & Query-focused summary      \\
			\hspace{0.5em} Snippet-abstractive-Llama & Query-focused summary        \\
			\midrule
			\multicolumn{2}{@{}l}{\textit{Reformulation}} \\
			\hspace{0.5em} Paraphrase-Gemma          & Paragraph rewriting          \\
			\hspace{0.5em} Paraphrase-Llama          & Paragraph rewriting              \\
			\hspace{0.5em} Propositions-Gemma        & Fact extraction                 \\
			\hspace{0.5em} Propositions-Llama        & Fact extraction                    \\
			\bottomrule
		\end{tabular}
	}
	\caption{The fourteen document representations evaluated, grouped by transformation type. The original prepared document serves as baseline. Suffixes \textit{-Gemma} and \textit{-Llama} denote outputs produced by Gemma 3 27B and Llama 3.3 70B respectively.}
	\label{tab:rep_properties}
\end{table}

\subsection{Document Representations}

Each transformation operated on the same prepared document.  Preparation involved two steps applied to each retrieved Wikipedia page: first, tail sections unlikely to contain answers (e.g., "References," "See Also," "External Links") were removed, reducing length and extraneous content at little cost to answer-bearing text; second, documents were truncated to 10,750 tokens (using the Llama 3 tokenizer) to reduce length variability and prevent a generator's capacity to handle very long contexts from confounding the representation comparison.   Retrieved pages varied in length from 13 to 124,439 tokens, with a mean of 3,851 and a 99th percentile of 26,409; documents exceeding this threshold were predominantly pages with large lists rather than dense prose, and 1,005 of the 11,955 documents (8.4\%) were truncated. 

We evaluate fourteen document representations. The first is the Original baseline, the prepared document with no transformation. The remaining thirteen are transformations grouped into three types based on how they transform the source: \textit{selection} methods extract spans from the source;   \textit{summarisation} methods rewrite the source into a shorter form; \textit{reformulation} methods restructure the source while preserving content. Table~\ref{tab:rep_properties} lists the representations within each type. Query-dependence and relative size are reported alongside accuracy in Table~\ref{tab:accuracy}.

\subsubsection{Selection Representations}

Selection representations retain a subset of the original document's tokens or sentences. Snippet-BM25 and snippet-cross-encoder produce short, query-dependent snippets similar to those shown by conventional search engines, by scoring all sentences in the document against the query to identify a target sentence. Snippet-BM25 scores using BM25, while snippet-cross-encoder uses BAAI/bge-reranker-base. A snippet is constructed by expanding outward from the target sentence, adding neighboring sentences until the snippet reaches 100 words, prefixed by the page title and section header to provide source context.

The remaining three selection methods accept a configurable retention target, set to 50\% in our experiments (indicated by the "-50" suffix). RECOMP \cite{xu_recomp_2024} scores sentences by query relevance using a pre-trained bi-encoder (trained on Natural Questions) and drops the least relevant sentences to reach the target. LLMLingua2 \cite{pan_llmlingua-2_2024} frames compression as token-level classification, using a trained XLM-RoBERTa model to label each token as preserve or discard, retaining tokens in their original order. Selective-context \cite{li_compressing_2023} ranks sentences by self-information (surprisal computed from a small causal language model) and prunes low-information sentences.

\subsubsection{Summarisation Representations}

Summarisation representations reduce document size by generating new, shorter text rather than selecting spans from the original. Two methods (summary and snippet-abstractive) were each run with two transformation models, yielding four representations. Both methods are LLM-based; implementation details, including the transformation models used, are given in \S\ref{subsec:llm-xf-details}.

Summary is query-independent: each section of the document was summarised independently, and the per-section summaries reassembled under their original headers, so the section structure is preserved by reassembly rather than by the model. No target length or compression ratio was imposed; summary length was determined by the model, and the size differences reported in \S\ref{subsec:rep-stats} arise from the transformation models themselves, since both received an identical instruction (Figure~\ref{fig:prompt-summary}). Gemma's summaries were on average over 50\% larger than Llama's (43.5\% vs 28.1\% of the original document). Sections were used as the unit of summarisation because individual paragraphs often lack sufficient context for a coherent summary.

Snippet-abstractive is query-dependent: the full document and query were provided to the LLM in a single pass, prompting it to produce a concise, answer-oriented summary (Figure~\ref{fig:prompt-snippet-abstractive}). The model received the whole document so that it could identify answer-relevant content with full context, rather than committing to relevance decisions section by section. Both models produced short snippets under the 100-word target, though Gemma's were on average twice the size of Llama's (3.4\% vs 1.7\% of the original document).

\subsubsection{Reformulation Representations}

Reformulation representations restructure the original document while preserving its information content, neither compressing nor extracting from it. Two methods (paraphrase and propositions) were each run with two transformation models, yielding four representations. Both methods are query-independent.

Paraphrase rewrites each paragraph of the document in different words while preserving its meaning (Figure~\ref{fig:prompt-paraphrase}), then reassembles the paraphrased paragraphs with the original title, section, and subsection headers intact.

Following the proposition decomposition of \citet{chen_dense_2024}, we decompose each section of the document into a bullet list of self-contained factual statements (Figure \ref{fig:prompt-propositions}),  which are then reassembled by section. Unlike paraphrase, propositions change from prose to list structure, but typically expand rather than reduce the original word count.

\subsubsection{Implementation details for LLM-based transformations.}
\label{subsec:llm-xf-details}

LLM-based transformations were produced using Llama 3.3 70B Instruct and Gemma 3 27B IT, both quantized to FP8 and served via vLLM \cite{kwon_efficient_2023}. Each transformation was run with both models to permit cross-family comparison.
Paraphrase, summary, and propositions follow a common decomposition strategy: the document is split into units, each unit is independently transformed, and the results are reassembled with the original document structure preserved. This focuses each LLM call on a single unit of text, prioritizing transformation quality over attempting to transform the entire document in one pass. The unit of decomposition differs by method: paraphrase operates per paragraph, while summary and propositions operate per section. Snippet-abstractive is the exception: it processes the full document and query in a single pass, since identifying answer-relevant content requires global document context.  Prompts for all LLM-based transformations are in Appendix~\ref{app:prompts-transform}.

\subsubsection{Representation Statistics}
\label{subsec:rep-stats}

Table~\ref{tab:accuracy} reports relative size along with accuracy results for each representation. Relative size is the mean word count post-transformation as a percentage of the original document; values below 100\% indicate compression and values above indicate expansion.  Extractive and abstractive snippets reduce documents to under 4\% of their original size.  Summaries retain roughly a quarter to a half of the original document's word count, whereas paraphrasing, on average, slightly expands word count.  Gemma's summaries are over 50\% larger than Llama's, and its abstractive snippets are twice the size, indicating meaningfully different compression behavior between the transformation models.  Paraphrase and propositions are the only representations that expand rather than compress the original.

\subsection{Answer Generation}

Four open-source, long-context generator models were selected for evaluation: Llama 3.1 8B Instruct, Mistral Nemo 12B, Gemma 3 12B IT, and Qwen 3.5 9B.   Llama and Gemma are from the same family as one of the two transformation LLMs (Llama 3.3 70B and Gemma 3 27B IT), enabling comparison of whether generators favor transformations from the same model family.  All models were run with temperature 0 for reproducibility and a maximum output length of 100 tokens.  Prompts for both closed-book and RAG can be found in Appendix~\ref{app:prompts-generator}.

\subsection{Evaluation}
\label{subsec:evaluation}

We evaluate generators on answer accuracy and computational cost, and representations on answer retention.

Answer accuracy was assessed using an LLM-judge (Qwen 2.5 32B), which classified each generated answer as correct or incorrect by comparing it to the gold answer(s). We departed from exact match, the standard KILT-NQ answer metric, because it credits an answer only if it matches an annotated string literally, penalising correct answers phrased differently --- the same gold-answer-set incompleteness that motivated our query filtering (\S\ref{subsec:dataset}).  We report accuracy as the proportion of queries judged correct, computed per generator and representation. The LLM-judge accuracy prompt can be found in Figure~\ref{fig:prompt-judge-accuracy}.

Answer retention measures whether a valid gold answer survives a transformation of the gold document, assessed by the same LLM-judge. This adapts the sufficient-context lens of \citet{joren_sufficient_2025} to a per-document setting: rather than asking whether the full retrieved context supports a plausible answer, we ask whether the gold document still supports its known answer after transformation. Measuring at the gold document gives a per-document signal of how the transformation affects answer-bearing content, which their measure does not isolate.  We report answer retention as the proportion of gold documents that support their known answers after transformation, computed per representation.  The LLM-judge retention prompt can be found in Figure~\ref{fig:prompt-judge-retention}.  Retention values are reported alongside accuracy in Table~\ref{tab:accuracy}.

To assess computational cost, we also measured query-time latency across a sample of 150 queries. For all representations, we measured the generator's time to first token (TTFT), which captures the cost of processing the input context and varies with representation size.   For query-dependent methods, query-time latency includes both transformation latency and the generator's TTFT, since both are incurred at query time.  Transformation latency was measured as the total sequential cost of transforming all retrieved documents for a query, although in practice, these transformations could be parallelized.  Because prefill cost scales quadratically with input length, a transformation that substantially reduces context size could offset its own cost through faster generation.  For query-independent methods, transformation is performed offline and its cost is excluded from the latency comparison.

%% file: sections/05_results.tex
\section{Results}
\label{sec:results}

We report results by the following research questions:

\begin{enumerate}[label=\textbf{RQ-\arabic*}]
	\item How do different representations of retrieved documents affect generator accuracy?
	\item  Do query-dependent representations outperform query-independent ones?
	\item Do generators prefer LLM-produced transformations?
\end{enumerate}

Although posed separately, the same factor recurs across all three questions: how much answer-bearing content each transformation preserves, which we focus on in \S\ref{sec:Discussion}.

All accuracy figures are LLM-judge accuracy on the KILT-NQ subset described in \S \ref{subsec:dataset}, and all comparisons are against each generator's own baseline on the original document unless stated otherwise.

% main accuracy table
\begin{table*}[t]
	\centering
	\small
	\setlength{\tabcolsep}{5pt}
	\sisetup{
		table-space-text-post = {\textsuperscript{$\blacktriangledown\blacktriangledown\blacktriangledown$}}
	}
	% Shorthand for significance markers
	\newcommand{\dnA}{\textsuperscript{$\blacktriangledown$}}
	\newcommand{\dnB}{\textsuperscript{$\blacktriangledown\blacktriangledown$}}
	\newcommand{\dnC}{\textsuperscript{$\blacktriangledown\blacktriangledown\blacktriangledown$}}
	\newcommand{\upA}{\textsuperscript{$\blacktriangle$}}
	\newcommand{\upB}{\textsuperscript{$\blacktriangle\blacktriangle$}}
	\newcommand{\upC}{\textsuperscript{$\blacktriangle\blacktriangle\blacktriangle$}}
	\begin{tabular}{
			>{\raggedright\arraybackslash}p{3.8cm}
			c
			S[table-format=3.1]
			S[table-format=3.1]
			S[table-format=2.1]
			S[table-format=2.1]
			S[table-format=2.1]
			S[table-format=2.1]
		}
		\toprule
		& & & & \multicolumn{4}{c}{\textbf{Generator}} \\
		\cmidrule(lr){5-8}
		\textbf{Representation} &
		\begin{tabular}[c]{@{}c@{}}\textbf{Query}\\\textbf{dep.}\end{tabular} &
		{\begin{tabular}[c]{@{}c@{}}\textbf{Relative}\\\textbf{size}\end{tabular}} &
		{\textbf{Retention}} &
		{\begin{tabular}[c]{@{}c@{}}\textbf{Qwen 3.5}\\\textbf{9B}\end{tabular}} &
		{\begin{tabular}[c]{@{}c@{}}\textbf{Gemma 3}\\\textbf{12B}\end{tabular}} &
		{\begin{tabular}[c]{@{}c@{}}\textbf{Mistral-Nemo}\\\textbf{12B}\end{tabular}} &
		{\begin{tabular}[c]{@{}c@{}}\textbf{Llama 3.1}\\\textbf{8B}\end{tabular}} \\
		\midrule
		Closed-book                       & & {--} & {--} & 40.0\dnC & 45.5\dnC & 47.5\dnC & 51.7\dnC \\
		Gold-only                         & & {--} & {--} & 87.6\upC & 85.5\upC & 84.8\upC & 84.0\upC \\
		Gold-5x                           & & {--} & {--} & 87.7\upC & 85.7\upC & 80.8\upC & 84.4\upC \\
		\midrule
		\rowcolor{gray!20}
		Original (baseline)               & & 100.0 & 98.6 & 81.3 & 80.6 & 73.9 & 80.0 \\
		\multicolumn{8}{l}{\textit{Selection}} \\
		\hspace{0.5em} Snippet-bm25                 & \checkmark & 3.6 & 52.8 & 60.7\dnC & 58.3\dnC & 60.3\dnC & 58.9\dnC \\
		\hspace{0.5em} Snippet-cross-encoder        & \checkmark & 3.5 & 74.3 & 68.3\dnC & 67.5\dnC & 67.8\dnC & 67.1\dnC \\
		\hspace{0.5em} Recomp-extractive-50         & \checkmark & 51.3 & 96.9 & 79.9\dnB & 79.3     & 76.0\upB & 78.3\dnA \\
		\hspace{0.5em} LLMLingua2-50                &            & 47.0 & 98.5 & 79.0\dnC & 76.4\dnC & 76.5\upB & 77.6\dnB \\
		\hspace{0.5em} Selective-context-50         &            & 51.3 & 81.9 & 69.8\dnC & 67.3\dnC & 71.0\dnB & 68.8\dnC \\
		\multicolumn{8}{l}{\textit{Summarisation}} \\
		\hspace{0.5em} Summary-Gemma                &            & 43.5 & 97.8 & 78.8\dnC & 79.1\dnA & 77.9\upC & 78.7     \\
		\hspace{0.5em} Summary-Llama                &            & 28.1 & 94.4 & 76.8\dnC & 76.8\dnC & 76.5\upB & 70.6\dnC \\
		\hspace{0.5em} Snippet-abstractive-Gemma    & \checkmark & 3.4 & 95.7 & 77.5\dnC & 76.4\dnC & 76.2\upA & 77.0\dnC \\
		\hspace{0.5em} Snippet-abstractive-Llama    & \checkmark & 1.7 & 95.2 & 80.8     & 78.3\dnA & 75.5     & 80.7     \\
		\multicolumn{8}{l}{\textit{Reformulation}} \\
		\hspace{0.5em} Paraphrase-Gemma             &            & 103.3 & 98.9 & 80.5     & 79.4     & 77.0\upC & 79.8     \\
		\hspace{0.5em} Paraphrase-Llama             &            & 105.3 & 99.2 & 81.2     & 80.7     & 74.8     & 79.8     \\
		\hspace{0.5em} Propositions-Gemma           &            & 119.0 & 98.8 & 81.1     & 80.0     & 73.6     & 80.3     \\
		\hspace{0.5em} Propositions-Llama           &            & 121.2 & 98.6 & 79.8\dnA & 80.0     & 72.4     & 80.1     \\
		\bottomrule
	\end{tabular}
	\caption{LLM-judge accuracy (\%) across document representations and generator models. \textbf{Query dep.} indicates whether the representation incorporates the query; \textbf{Relative size} (\%) is mean word count as a percentage of the original document (omitted for the closed-book and gold-only/gold-5x reference conditions); \textbf{Retention} (\%) is the proportion of gold documents in which a valid gold answer survives transformation (\S\ref{subsec:evaluation}). The shaded row marks the \textit{original (baseline)} for each generator. Suffixes \textit{-Gemma} and \textit{-Llama} denote representations produced by Gemma 3 27B and Llama 3.3 70B respectively. Markers indicate significance vs.\ Original (McNemar's test, $n{=}2{,}391$): $\blacktriangle$/$\blacktriangledown$ denote performance above/below baseline, and the number of markers reflects the significance level (one: $p<0.05$, two: $p<0.01$, three: $p<0.001$). Unmarked cells are not significantly different from baseline.}
	\label{tab:accuracy}
\end{table*}

\subsection{RQ-1}
\label{subsec:rq1}

\textit{How do different representations of retrieved documents affect generator accuracy?}

% summary of results and findings
Three of the four generators --- Qwen, Gemma, and Llama --- behave consistently; Mistral-Nemo does not, and is treated separately below. For these three generators, two findings stand out. First, no representation significantly improves on the baseline; the closest --- snippet-abstractive-Llama and the reformulation methods --- match it but none exceeds it (Table~\ref{tab:accuracy}). Second, accuracy tracks gold-answer retention. When retention is high, the other properties of a representation --- its wording, structure, and length --- have limited effect; representations compressed to under 4\% of the original range from baseline-equivalent to more than 20 points below, depending on what they retain rather than how much. This pattern holds regardless of whether a representation incorporates the query; whether query-dependence helps on balance is taken up in \S\ref{subsec:rq2}.

% accuracy vs retention figure
\begin{figure}[t]
	\centering
	\includegraphics[width=\columnwidth]{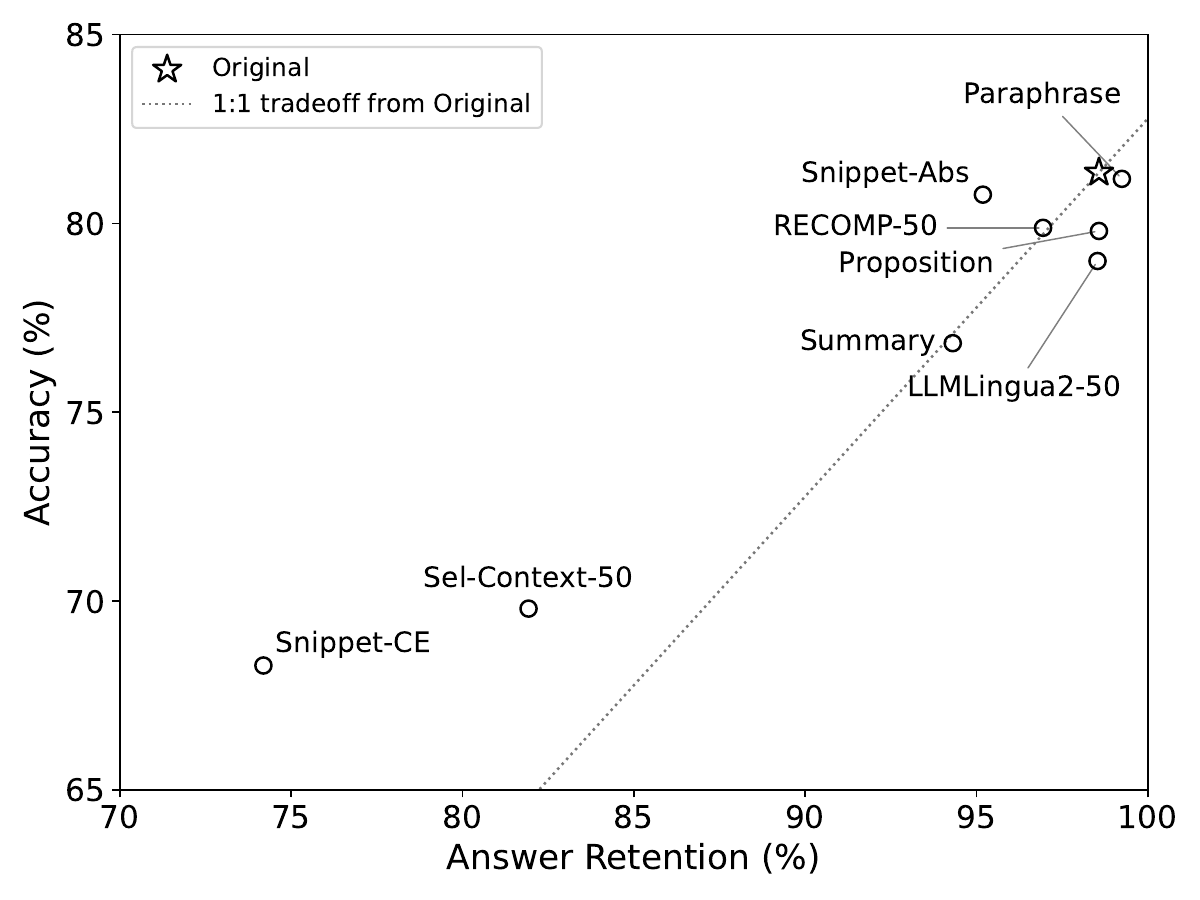}
	\caption{Answer accuracy versus answer retention for each document representation, shown for the Qwen-3.5-9B generator with Llama-3.3-70B as the transformation model. The dotted line has slope 1 and passes through the original document, so along it accuracy and retention fall by equal amounts relative to the original. A representation above the line gives up less accuracy than retention relative to the original, while one below it gives up more.  Snippet-BM25 is omitted as its 52.8\% retention falls below the displayed range.}
	\label{fig:accuracy_v_retention}
\end{figure}

% explanation of main finding
The relationship between answer retention and accuracy is shown in Figure~\ref{fig:accuracy_v_retention}. The high-retention methods --- those preserving the gold answer in roughly 95\% of documents or more --- cluster near the original and maintain accuracy close to baseline, sitting close to the slope-1 line; departures are small in both directions and, where statistically significant (recomp-extractive-50, LLMLingua2-50), amount to only 1--2 accuracy points (Table~\ref{tab:accuracy}). The methods with substantial retention loss (selective-context-50, snippet-cross-encoder) sit well above the line, losing markedly less accuracy than retention; snippet-BM25, omitted from the figure, is the most extreme case --- its retention falls to 52.8\% while accuracy holds at 60.7\% (Table~\ref{tab:accuracy}). Across the representations, then, accuracy loss is smaller than retention loss alone would predict --- a gap to which several mechanisms not captured by gold-answer retention may contribute. One is parametric knowledge from the generator. Another is answer content surviving in non-gold retrieved documents --- a manual review confirmed this directly, finding cases where non-gold documents in the retrieved set contained the gold answer. The first echoes Joren et al. (2025), who document that generators answer correctly via parametric knowledge even when the context does not support the answer; the second is specific to our design --- because we measure retention on the gold document alone rather than over the full retrieved context as they do, answer content in a non-gold document contributes to accuracy without registering as retention. Thus retention measured on the gold document is a lower bound on the generator's actual access to answer-bearing content.

\begin{figure*}[t]
	\centering
	\makebox[0.48\textwidth]{(a) Original}\hfill\makebox[0.48\textwidth]{(b) Snippet-abstractive}
	\includegraphics[width=\textwidth]{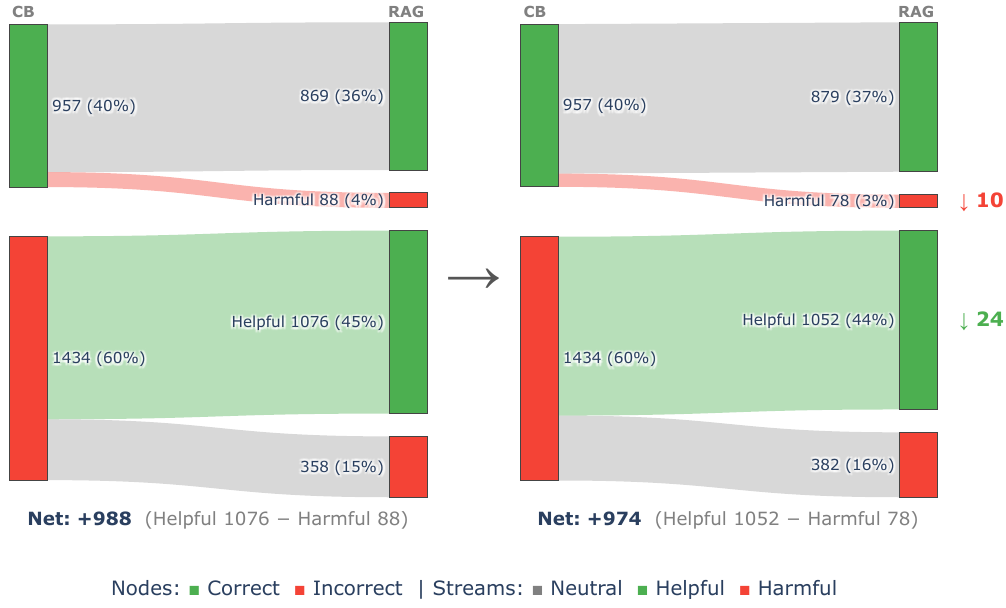}
	\par\smallskip
	
	\caption{Per-query transitions between closed-book (CB) and RAG outcomes for Qwen 3.5 9B, under two RAG conditions: (a) the original retrieved documents and (b) snippet-abstractive (Llama). Green nodes denote correct answers, red nodes denote incorrect answers. The four flows correspond to the categories defined in \S\ref{subsec:rq1}: \textit{helpful} (CB incorrect $\rightarrow$ RAG correct), \textit{harmful} (CB correct $\rightarrow$ RAG incorrect), and two neutral flows where outcomes agree. Net is the per-query gain over closed-book, computed as helpful minus harmful queries. Moving from (a) to (b), snippet-abstractive reduces harmful queries by 10 (88 $\rightarrow$ 78) but reduces helpful queries by 24 (1076 $\rightarrow$ 1052); because the helpful loss is larger, net gain falls from +988 to +974 and overall accuracy sits below the original.}
	\label{fig:sankey}
\end{figure*}

% how important is representation?
Four representations illustrate that representation has limited effect on accuracy when retention is high. Paraphrase isolates the effect of wording: it rewords every paragraph while leaving content, length, and retention roughly constant (103\% length, 99.2\% retention). Propositions tests a structural change as well, reformulating prose into bullet-lists of factual statements (119\% length, 98.8\% retention). Snippet-abstractive-Llama adds radical compression, generating an answer-oriented summary at a fraction of the original length (2\% length, 95.2\% retention). Despite very different representations, all three perform close to the baseline across generators. LLMLingua2-50 goes further, pruning tokens to produce incoherent text (47\% length, 98.5\% retention) --- yet accuracy lands within 1–4 points of paraphrase.

% does size predict accuracy?
Methods that compress to similar sizes can differ sharply in accuracy depending on what they retain. Snippet-abstractive-Llama, snippet-BM25, and snippet-cross-encoder all reduce documents to under 4\% of their original word count. Yet snippet-abstractive-Llama matches the baseline to within 2.3 points across all four generators, while snippet-BM25 and snippet-cross-encoder lose 13–22 points across Qwen, Gemma, and Llama (McNemar's test, p < 0.001). The methods differ in how they identify answer-relevant content: snippet-abstractive uses a 70B LLM with access to the full document and query, while the others score sentences against the query in isolation. The accuracy gap is attributable to retention: snippet-abstractive-Llama preserves the gold answer in 95\% of documents, while snippet-BM25 and snippet-cross-encoder retain it in only 52.8\% and 74.3\% (Table~\ref{tab:accuracy}).

% mistral is an outlier
Mistral-Nemo behaves differently from the other three generators. Its accuracy on the original retrieved set is 73.9\%, 6–7 points below Qwen, Gemma, and Llama. This deficit is not explained by parametric knowledge or single-document handling: closed-book accuracy (47.5\%) is the second-highest of the four generators, and gold-only accuracy (84.8\%) is mid-range. The deficit appears specifically when Mistral is given the full retrieved set.

Two findings point to factors beyond retention. First, gold-5x scores 4 points below gold-only.  Since the two conditions contain the same content, the drop must come from length, repetition, or both.  Second, Mistral's accuracy among high-retention representations does not track retention in the way it does for the other generators. Reformulation methods preserve retention at the highest rates but, with the exception of paraphrase-Gemma, are not significantly different from baseline for Mistral. Summarisation methods retain fewer gold answers and tend to reduce accuracy for the other three generators, yet improve Mistral's results. For Mistral, retention appears to matter less and length appears to matter more. Reformulation methods, which preserve both retention and length, perform roughly at baseline; summarisation methods, which reduce length at some cost to retention, improve performance.

% helpful/harmful zoom in
To examine the per-query impact, Figure~\ref{fig:sankey} decomposes accuracy into transitions between closed-book and RAG conditions. Each query falls into one of four categories: \textit{helpful} (closed-book incorrect, RAG correct), \textit{harmful} (closed-book correct, RAG incorrect), or one of two neutral categories where the RAG and closed-book outcomes agree. The accuracy gain from RAG over closed-book is the difference between the helpful and harmful rates: a representation improves on closed-book only when the helpful rate exceeds the harmful rate.  For the baseline, 33\textendash 45\% of queries are helpful, while 4\textendash 7\% are harmful. Snippet-abstractive (Llama) achieves a comparable helpful rate (34\textendash 44\%) but has the lowest harmful rate across most generators (3\textendash 6\%), suggesting it introduces less noise into queries the generator could already answer. Despite the lower harmful rate, snippet-abstractive's overall accuracy does not exceed the baseline: its advantage on harmful queries (a lower harmful rate) is cancelled by a roughly equal disadvantage on helpful queries (a lower helpful rate).

\subsection{RQ-2}
\label{subsec:rq2}

\textit{Do query-dependent representations outperform query-independent ones?}

\subsubsection{Answer Accuracy}

% overview
Query-dependent representations do not systematically outperform query-independent ones. Across Qwen, Gemma, and Llama, neither the strongest query-dependent method (snippet-abstractive-Llama) nor the strongest query-independent methods (paraphrase-Gemma, paraphrase-Llama, propositions-Gemma) significantly differ from the baseline (Table \ref{tab:accuracy}). Incorporating the query into the representation does not, on its own, improve answer accuracy. A per-query breakdown of the strongest query-dependent method points to why this may be. On Qwen, snippet-abstractive-Llama turns fewer correct closed-book answers wrong than the original (correct $\rightarrow$ incorrect) but also rescues fewer wrong ones (incorrect $\rightarrow$ correct); because the lost help (24 queries) exceeds the avoided harm (10 queries), it ends up just below baseline rather than above it, though not significantly so (Figure~\ref{fig:sankey}). This is a single method on a single generator, but it suggests query-dependent tailoring may trade coverage for safety: discarding non-answer content removes distracting noise but also removes content the generator could otherwise have used.

\subsubsection{Query-time Latency}

\begin{figure*}[t]
	\centering
	\includegraphics[width=0.9\textwidth]{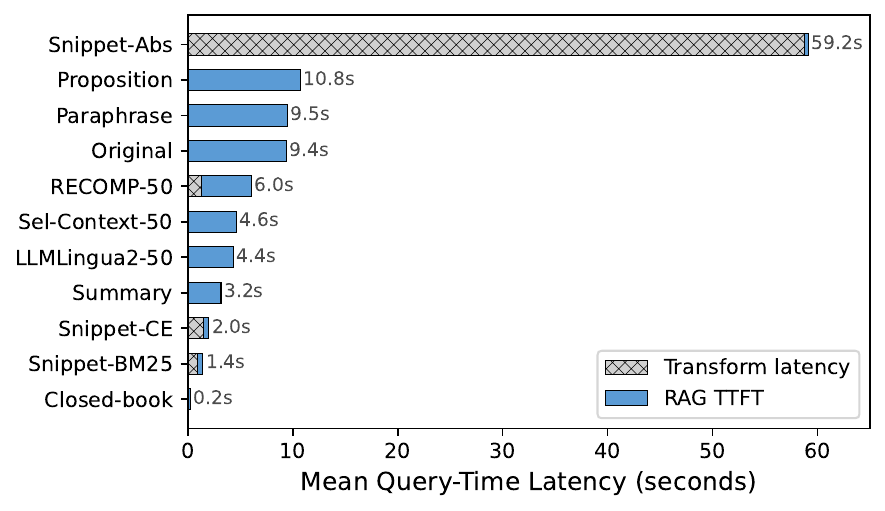}
	\caption{Mean query-time latency per representation, decomposed into transformation latency (hatched) and generator time-to-first-token (solid). Transform latency values are averaged across the two transformation models and TTFT values are averaged across the four generator models. Query-independent methods  incur no query-time transformation cost since they are pre-computed; their bars are identical across panels. Query-dependent methods incur transformation cost at query time, shown as the hatched component.}
	\label{fig:qt_bar}
\end{figure*}

% latency landscape
Query-time latency varies substantially across representations. Figure \ref{fig:qt_bar} decomposes  query-time latency into transformation latency (incurred only for query-dependent methods) and the generator's time-to-first-token (TTFT), which scales with input length. The original document's latency is approximately 9 seconds, entirely TTFT. Reducing input size lowers it proportionally: summary and LLMLingua2-50 roughly halve it, while extractive snippets cut it by over 75\%. Query-dependent methods add transformation latency on top, ranging from roughly 1 second for RECOMP-extractive-50 to 59 seconds for snippet-abstractive, whose query-time latency is almost entirely transformation cost, since the transformation model must process the full document and query before generating the snippet. Query-independent transformations, by contrast, can be computed ahead of time and stored, leaving only TTFT at query time.

% abstractive scaling cost benefit 
\begin{figure}[H]
	\centering
	
	\begin{subfigure}{\columnwidth}
		\centering
		\includegraphics[width=\columnwidth]{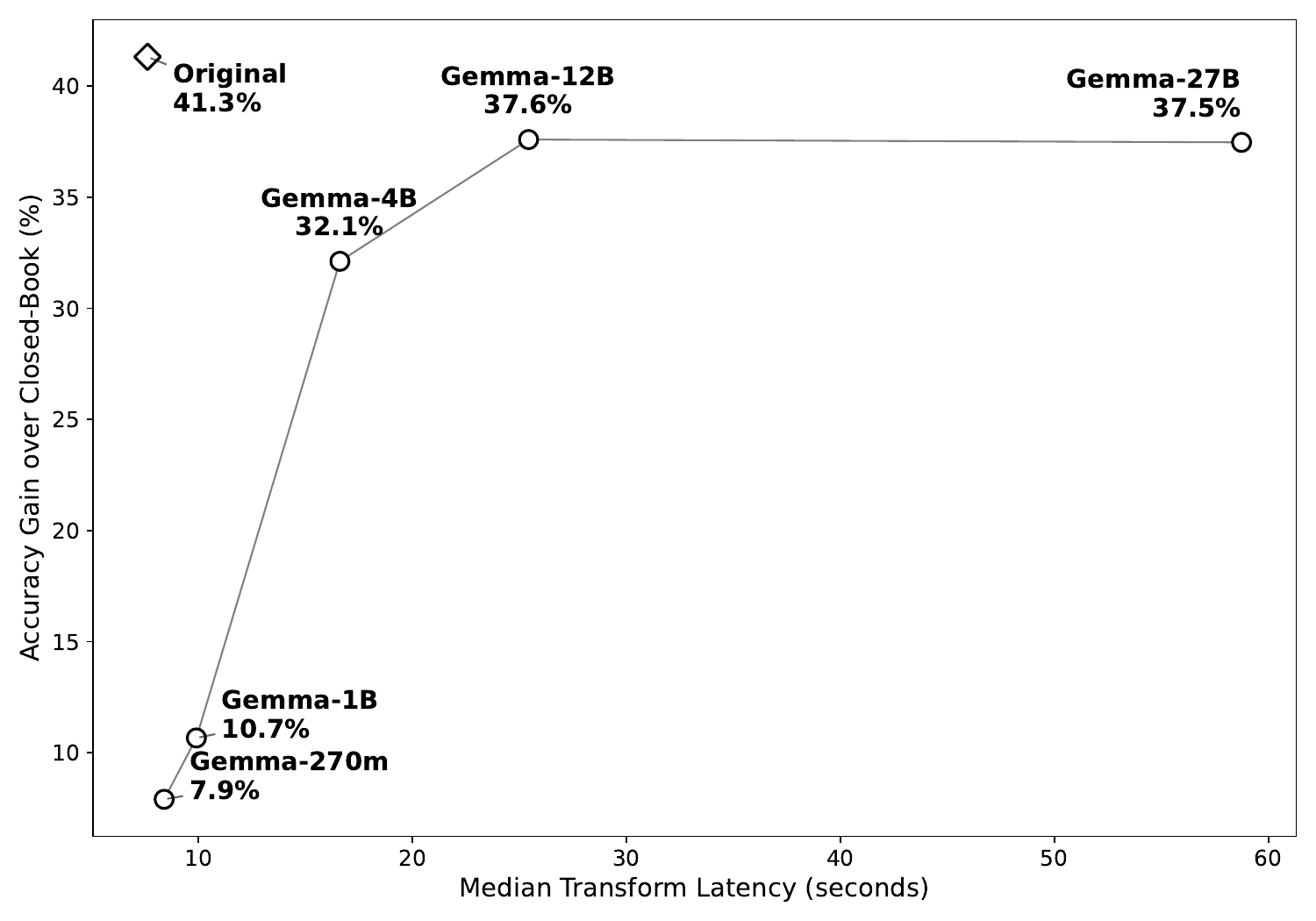}
		\caption{}
		\label{fig:snippet-scaling-a}
	\end{subfigure}

	\vspace{0.5em}

	\begin{subfigure}{\columnwidth}
		\centering
		\includegraphics[width=\columnwidth]{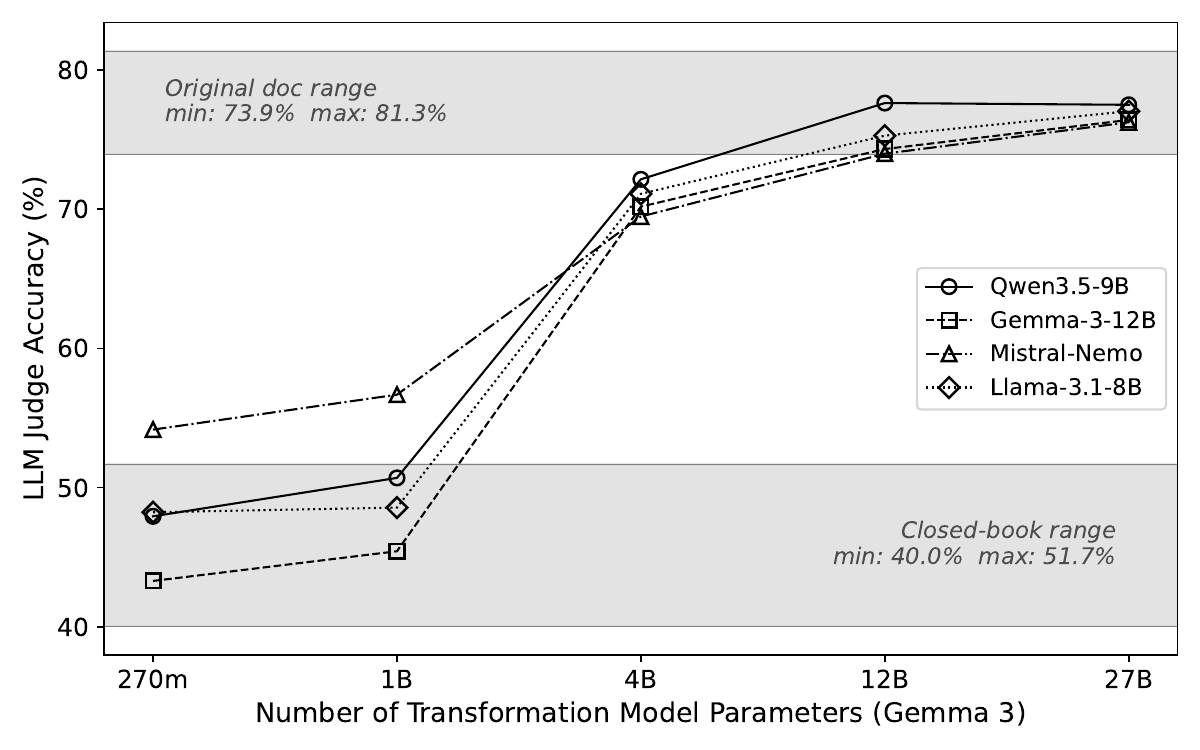}
		\caption{}
		\label{fig:snippet-scaling-b}
	\end{subfigure}
	
	\caption{Snippet-abstractive performance as the Gemma 3 transformation model is scaled from 270M to 27B parameters. (a) Accuracy gain over closed-book versus median transformation latency for Qwen-3.5-9B; each circle denotes a transformation model size and the diamond denotes the original document baseline (no transformation). (b) LLM-judge accuracy versus transformation-model size for all four generators, with the original-document and closed-book accuracy ranges shown as shaded bands. Accuracy improves steeply from 270M to 4B and plateaus thereafter: Gemma-12B matches Gemma-27B accuracy at roughly half the transformation latency. Even at 12B, snippet-abstractive remains slower than query-independent alternatives that achieve comparable accuracy (Figure~\ref{fig:accuracy_v_qt}).}
	\label{fig:snippet-scaling}
\end{figure}

\subsubsection{The accuracy-latency frontier}

\begin{figure*}[t]
	\centering
	\includegraphics[width=0.8\textwidth]{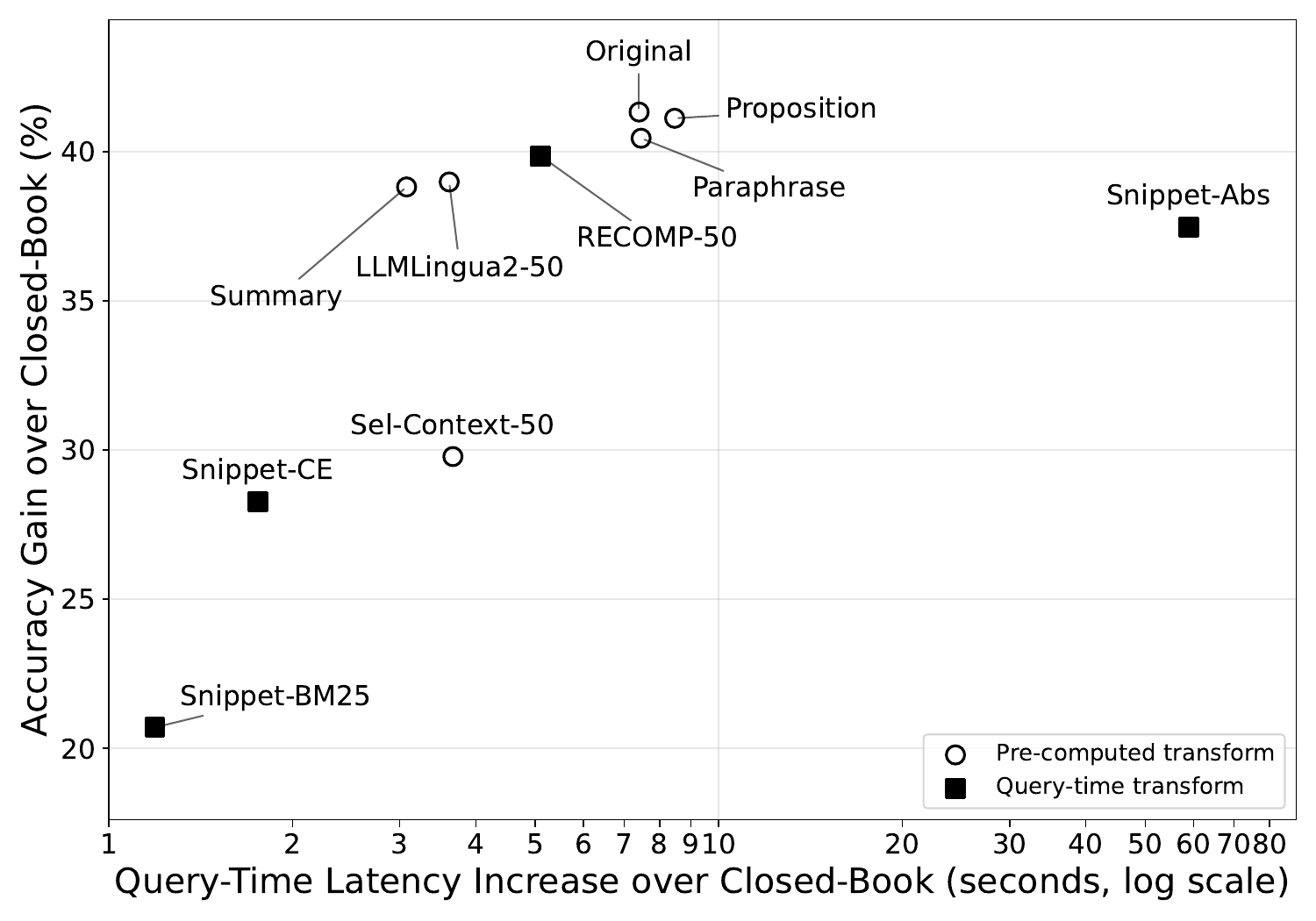}
	\caption{Accuracy gain over closed-book vs. query-time latency increase over closed-book, for Qwen 3.5 9B with Gemma 3 27B as the transformation model. Query-time latency is plotted on a log scale. Circles denote pre-computed (query-independent) representations, whose transformation cost is incurred offline; squares denote query-time (query-dependent) representations, whose transformation cost is incurred at query-time.}
	\label{fig:accuracy_v_qt}
\end{figure*}

% accuracy-latency frontier
Figure \ref{fig:accuracy_v_qt} plots accuracy gain over closed-book against query-time latency to characterise the accuracy-latency tradeoff faced when selecting a representation. The ideal representation would sit in the top-left corner: high accuracy at low latency. In our results, no representation occupies that corner; faster methods generally come at the expense of accuracy. The original document and reformulation methods are among the most accurate but also the slowest. Two query-independent representations, summary and LLMLingua2-50, offer a favourable tradeoff, retaining most of the baseline's accuracy gain at roughly a third of the latency. RECOMP-extractive-50 reaches similar accuracy but at higher latency, due to its query-time transformation overhead. The sentence-scored snippets are the fastest representations but operate at a substantially lower accuracy. Snippet-abstractive matches the baseline accuracy on three of four generators, but at roughly six times the latency; whether this is justified depends on the deployment's latency budget, and \S \ref{subsec:sa-scaling} examines whether a smaller transformation model can close the gap.

% synthesis
Taken together, the accuracy and latency results indicate that query-dependence does not, in general, justify its overhead. The query-dependent method with the strongest accuracy (snippet-abstractive-Llama) matches the baseline accuracy on three of four generators but at six times the query-time latency, while the query-dependent methods with low latency (snippet-BM25, snippet-cross-encoder) achieve substantially lower accuracy than their query-independent counterparts. Where a query-time transformation is required, snippet-abstractive is the strongest option, and \S\ref{subsec:sa-scaling} examines whether a smaller transformation model can reduce its latency without sacrificing accuracy.  More broadly, no single representation dominates across the accuracy-latency frontier; the right choice depends on the deployment's latency budget.

\subsubsection{Reducing Snippet-abstractive Latency}
\label{subsec:sa-scaling}

The accuracy-latency analysis in \S \ref{subsec:rq2} identified snippet-abstractive as a high-latency outlier: it matches the baseline accuracy but at roughly six times the query-time latency, almost entirely due to the document transformation. Since this latency scales with the size of the transformation model, a natural question is whether a smaller transformation model can produce acceptably effective snippets at a lower cost. Figure~\ref{fig:snippet-scaling} reports accuracy and latency as the Gemma 3 transformation model is scaled from 270M to 27B parameters. Accuracy improves steeply from 270M to 4B, then plateaus. The 270M and 1B models produce snippets that perform barely above the closed-book accuracy (Qwen accuracy of 47\% and 51\% respectively, against closed-book of 40\% and a 27B-snippet of 77.5\%). At 4B, accuracy reaches 71\% on Qwen, and at 12B it reaches 77.6\%, essentially identical to the 27B model's 77.5\%. Transformation latency at 12B is approximately half that of 27B.

Smaller transformation models can therefore reduce snippet-abstractive's transformation latency by roughly half without sacrificing accuracy, but cannot eliminate it. The 12B model is the smallest that matches 27B accuracy; below that, accuracy drops sharply. And even at 12B, snippet-abstractive must be computed at query time, and remains slower than query-independent alternatives (summary-Gemma, LLMLingua2-50) that achieve comparable accuracy. The frontier conclusion from \S \ref{subsec:rq2} is unchanged: pre-computed transformations remain the strongest option at this accuracy tier.

% qt bars

% accuracy vs qt

\subsection{RQ-3}
\label{subsec:rq3}

\textit{Do generators prefer LLM-produced transformations?}

We first ask whether generators favour LLM-produced representations over non-LLM ones. At matched retention, the LLM/non-LLM distinction does not predict accuracy. Non-LLM methods with high retention (RECOMP-extractive-50, 96.9\%; LLMLingua2-50, 98.5\%) perform within 1-2 points of LLM-produced summary-Gemma (97.8\% retention) on the three retention-tracking generators. Non-LLM methods that underperform (snippet-BM25, snippet-cross-encoder, selective-context-50) do so because they discard answer-bearing content (52.8\%, 74.3\%, and 81.9\% retention respectively). 

\begin{figure*}[t]
	\centering
	\includegraphics[width=\textwidth]{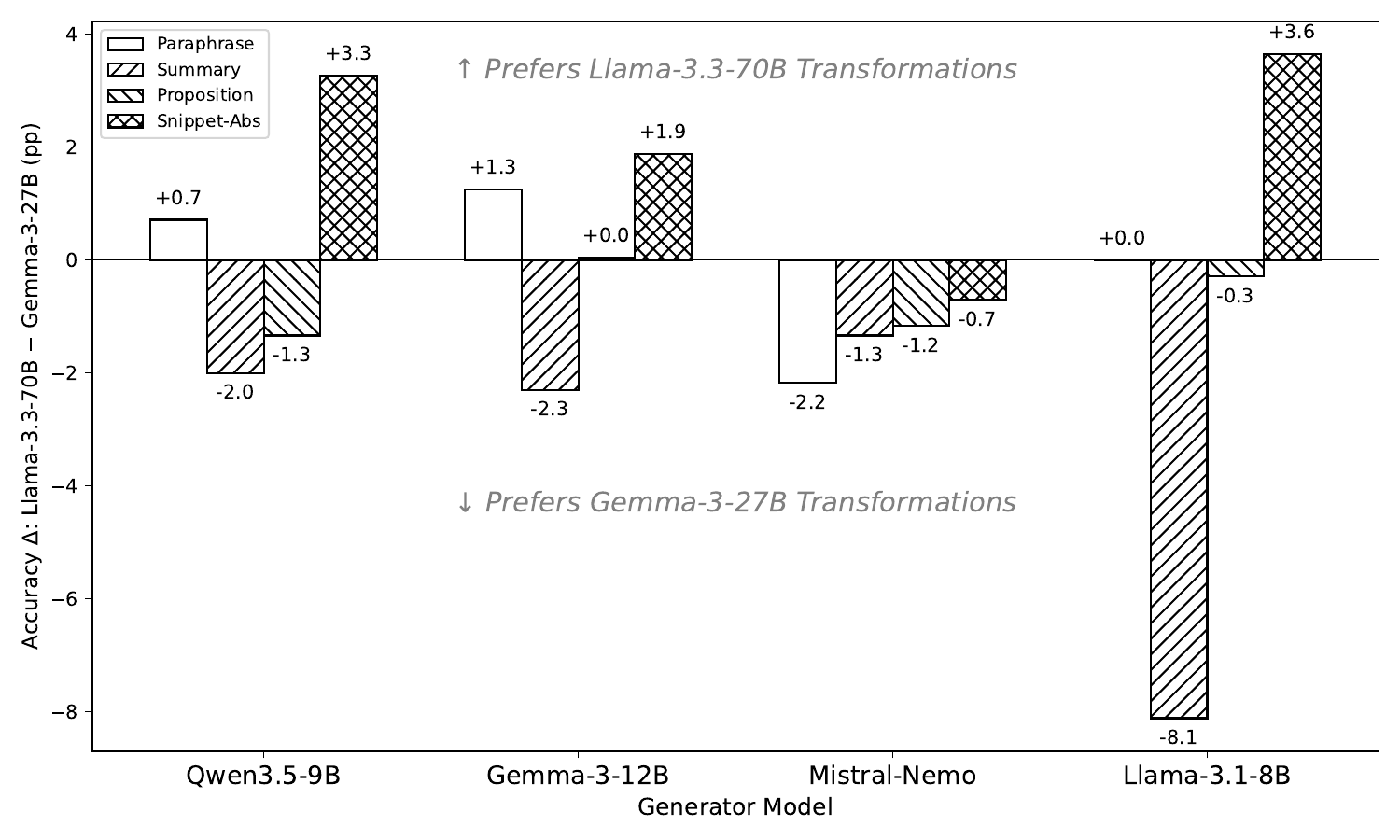}
	\caption{Generator family bias: Does Gemma-12B favor gemma 27B transforms? "+" favors Llama-3.3-70B, "-" favors Gemma-3 27B.}
	\label{fig:family_bias}
\end{figure*}

We next ask whether generators prefer transformations from their own model family. Figure \ref{fig:family_bias}  reports each generator's preference between representations produced by Gemma-3-27B and Llama-3.3-70B. If family bias were present, Gemma-3-12B would consistently prefer Gemma-3-27B representations and Llama-3.1-8B would prefer Llama-3.3-70B representations. This pattern does not appear. For summary transformations, all four generators prefer Gemma representations, including Qwen and Mistral, which share no family relationship with either transformer; this is consistent with Gemma summaries retaining the gold answer more often than Llama summaries (97.8\% vs 94.4\%). For snippet-abstractive, retention is closely matched (95.7\% vs 95.2\%), so the Llama preference shown by three of four generators reflects a property of the representations we do not isolate. It is not, however, family alignment: non-family generators (Qwen, Mistral) show the same preference. Paraphrase and propositions show no consistent preference.  Across both comparisons, neither the LLM/non-LLM distinction nor model family predicts accuracy independent of retention. 

% family bias

%% file: sections/06_discussion.tex
\section{Discussion}
\label{sec:Discussion}

\subsection{Answer retention as the dominant factor}
\label{sec:subsec-d1}

Across the fourteen representations evaluated, the strongest predictor of generator accuracy is whether the gold answer survives the transformation (answer retention), not the representation in which it survives.  The retention measure adapts the sufficient-context lens of \citet{joren_sufficient_2025} to a per-document, post-transformation setting: rather than asking whether the full retrieved context supports a plausible answer, we ask whether the gold document still supports a known answer after transformation. Transformation methods that retain the answer at high rates cluster near the baseline accuracy; methods that lose answer-bearing content suffer accuracy drops that scale with what they discard. This pattern holds across selection, summarisation, and reformulation methods, across LLM-produced and non-LLM transformations, and across three of the four generators (Mistral-Nemo, discussed in \S\ref{sec:Limitations}, is the exception). Representational dimensions appear secondary: paraphrase alters wording, propositions alter structure, and LLMLingua2-50 disrupts coherence, yet all three perform near baseline when retention is high. What these methods share is high answer retention; what they differ in does not predict accuracy among them.

\subsection{Reinterpreting prior work through retention}
\label{sec:subsec-d2}

A natural objection is that the retention finding is obvious: of course generators need the answer to be present. The non-obvious claim concerns attribution: recent methods report accuracy effects and credit them to a proposed mechanism, but retention --- how well each mechanism happens to preserve answer-bearing content --- may account for more of those effects than the mechanism-based framing implies.

Recent work on RAG representation has explored methods aimed at improving generator accuracy, including hierarchical abstraction \cite{sarthi_raptor_2024}, proposition-level indexing \cite{chen_dense_2024}, and learned document rewriting \cite{kim_relevance_2025, li_align_2026}, alongside compression methods that aim to preserve accuracy under reduced input size \cite{xu_recomp_2024, pan_llmlingua-2_2024, li_compressing_2023}. Each varies one aspect of how retrieved content is represented and reports an accuracy effect (a gain, or preservation under compression), which is then attributed to the proposed mechanism. Our results raise the possibility that these effects are partly explained by retention. A retention-matched comparison in our data is consistent with this reading: RECOMP-extractive-50, LLMLingua2-50, and summary-Gemma — three methods that operate on the document via fundamentally different strategies (\S \ref{sec:experimental_setup}), yet retain the gold answer at comparable rates (~97–99\%) and perform within a narrow accuracy band on three of four generators.

LLMLingua2-50 makes the point most clearly.  Its compression objective is framed around faithfulness, but faithfulness here means only that the output is an extractive subsequence of the input text. This fixes what kind of text is kept, not whether the answer is among it: two equally faithful compressions of the same document can differ entirely in retention depending on which sentences they drop.  The text LLMLingua2-50 produces is incoherent (\S\ref{subsec:rq1}) and the method is query-independent, so it tailors its output neither to readability nor to the question. Yet in our setup neither limitation impairs it, retaining the gold answer in 98.5\% of documents and landing within roughly two points of baseline on Qwen. Because the method is extractive by construction, faithfulness is a property every output shares equally, and a property that does not vary across representations cannot explain why their accuracy varies.  Retention does vary, and accuracy with it. We do not claim this re-attributes the results \citet{pan_llmlingua-2_2024} report; we observe only that, in our data, the faithfulness they emphasise is constant across its outputs, while the accuracy it yields is not.

A parallel argument applies to work that characterises which properties of retrieved content drive generation. Utility \cite{tian_is_2025}, generator-specific utility \cite{zhang_llm-specific_2025}, and logical connectivity across passages \cite{chang_what_2025} are each reported as properties to which generators are sensitive, without retention measured alongside. Our results cannot adjudicate these properties, but they show that retention alone reproduces the accuracy patterns these properties are invoked to explain: methods spanning selection, summarisation, and reformulation cluster within a few points of baseline when retention is held high, and drop in proportion to what they discard when it is not. These properties divide by whether they concern answer-bearing content. For those that do --- utility, generator-specific utility, and logical connectivity --- retention is a rival explanation, since an effect attributed to the property could instead reflect how well the method preserved the answer. The distraction effect \citep{cuconasu_power_2024, amiraz_distracting_2025} is the exception: it concerns content that is not answer-bearing, which retention does not measure, so the two are genuinely separate.

This is not a claim that the proposed mechanisms or properties are wrong; relevance scoring, surprisal-based pruning, hierarchical abstraction, and learned-document rewriting may each contribute beyond retention, as may utility and logical connectivity, in ways our experiment cannot detect. The claim is methodological: absent experiments that hold retention constant and vary the mechanism or property independently, the field cannot distinguish "this method works because of mechanism X" from "this method works because it happens to preserve answers," nor "generators are sensitive to property Y" from "generators are sensitive to whether the answer survives, and Y correlates with that." Our own design has the same limitation: retention varies as a consequence of method rather than being controlled.

\subsection{ Source effects: human-vs-LLM and family preference}

Prior work finds that LLMs favour LLM-produced content --- choosing it over human-authored text \cite{laurito_aiai_2025}, and favouring their own generations over other models' in evaluation \cite{panickssery_llm_2024}. Within RAG the picture is mixed: \citet{tan_blinded_2024} find generators favour generated context over retrieved even when it is incorrect, while \citet{chen_llms_2025} find factual accuracy overrides this preference in fact-centric settings. Whether either preference extends to the representation of retrieved documents has been open.

In our setup, retention accounts for what looks like a source effect in both the LLM-versus-non-LLM and family-alignment cases. At matched retention, the LLM/non-LLM distinction does not predict accuracy: high-retention non-LLM methods perform within 1–2 points of LLM-produced summary-Gemma on the three retention-tracking generators, and the non-LLM methods that underperform do so because they discard answer-bearing content (\S\ref{subsec:rq3}). Family alignment shows the same pattern. Where a generator appears to prefer same-family transformations, non-family generators show the same preference, and the apparent effect tracks either retention differences between transformation models or unidentified properties shared across all generators.

This result extends \citet{chen_llms_2025}, who found that the factual accuracy of the content, rather than whether it is human- or LLM-produced, drives generator output, from a setting where source is varied by choosing between generated and retrieved contexts to one where it is varied by transforming the same retrieved document with two different LLMs.  In both, the apparent source effect dissolves once a more proximal property of the content is controlled --- factual accuracy in their case, answer retention in ours.

\subsection{Why query-dependence underdelivers}
\label{sec:subsec-d3}

Query-dependent representations might be expected to improve generator accuracy by tailoring retrieved content to the question. None does (Section \ref{subsec:rq2}). The retention pattern (\S \ref{sec:subsec-d1}) accounts for the spread of results.

Aggressive query-dependent methods compress to 3–4\% of the original content length. At this budget, retention depends on how well the method identifies answer-relevant content. The two LLM-based variants do this well: snippet-abstractive-Llama and snippet-abstractive-Gemma retain the gold answer in 95.2\% and 95.7\% of cases respectively and perform within 3 points of the baseline on three of four generators,  though at six times the query-time latency, almost entirely transformation overhead. Snippet-BM25 and snippet-cross-encoder score sentences against the query in isolation, retain the answer in only 52.8–74.3\% of cases, and lose 13–22 points. RECOMP-extractive-50 keeps 51\% of the document's sentences, leaving more room for the answer to survive imperfect relevance judgments. It preserves the answer at 96.9\%, though its bi-encoder is trained on NQ and its retention advantage may not transfer to other query distributions.

Across these methods, query-awareness adds nothing beyond retention. Where retention is preserved, accuracy follows; where it is not, accuracy drops. The query-dependent framing does not predict performance independently of how much the method retains.

\subsection{Relation to prompt-sensitivity findings}

Our results may appear to conflict with work showing that language models are highly sensitive to surface features of their input.  Small changes to prompt formatting (e.g. separators, capitalisation, spacing) can cause large changes to task accuracy \cite{sclar_quantifying_2024}. Our manipulations of retrieved content are far more aggressive --- paraphrase rewrites every paragraph, propositions convert prose to bullet lists, LLMLingua2 prunes coherence at the token level --- yet none shifts accuracy by more than a few points when the answer is preserved. 

Several factors could explain the difference, including task type, and the different roles of instructions and retrieved content within a prompt. Our results speak to how the wording and structure of retrieved content affect accuracy when the answer is preserved; whether the same holds for instruction formatting remains open.

%% file: sections/07_conclusion.tex
\section{Conclusion}

We presented a controlled comparison of fourteen document representations --- an original baseline and thirteen transformations spanning selection, summarisation, and reformulation, in both query-dependent and query-independent variants --- evaluated across four open generators. By holding retrieval fixed and injecting answer-bearing content when needed, we isolated the effect of representation from that of retrieval, and we introduced answer retention, a per-document adaptation of the sufficient-context lens, to measure whether a transformation preserves the content needed to answer. Across these representations, answer retention is the primary determinant of generator accuracy: when retention is high, the wording, structure, and length of a representation have limited effect (RQ-1). Query-dependent representations do not systematically outperform query-independent ones, and query-dependence does not, in general, justify its overhead (RQ-2).  Generators show no preference for LLM-produced transformations, nor for transformations from their own model family, once retention is accounted for (RQ-3).

The representational dimensions tested here --- wording, structure, length, query-relevance, and whether content was produced by an LLM --- are characteristic targets of work designed around human readers. Our results show that, when answer retention is held constant, these dimensions together account for little of the variance in generator accuracy. We do not claim that representation is irrelevant in principle, only that the particular dimensions our transformations vary are not, in this setting, the ones that determine whether a generator can use a retrieved document.

The broader implication is methodological. Recent work attributes accuracy effects to the specific mechanism each method proposes, but how well that mechanism preserves answer-bearing content may explain more of those effects than the mechanism itself. Absent experiments that hold retention constant while varying the mechanism independently, an effect attributed to a mechanism cannot be distinguished from the retention that mechanism happens to produce.  Isolating the effect of representational dimensions across the full range of retention will require representations matched on retention but differing in representational dimensions.  Whether the finding extends to non-textual representations or to multi-hop tasks, where per-document retention is no longer well-defined, is a further open question.

%% file: sections/08_limitations.tex
\section{Limitations}
\label{sec:Limitations}

Our findings are subject to limitations of four kinds: the design isolates neither representation from retention (\S\ref{sec:lim-confound}) nor its dimensions from one another (\S\ref{sec:lim-decomp}); the study covers a narrow slice of representations, datasets, tasks, and models (\S\ref{sec:lim-scope}); it characterises representation only under idealised retrieval (\S\ref{sec:lim-goldinjection}); and the retention account, while dominant, is not the only factor at work (\S\ref{sec:subsec-d4}).

\subsection{Representational dimensions and retention are not independently varied}
\label{sec:lim-confound}

Retention is not a controlled variable in our design: each representation's retention is a byproduct of its transformation, not a parameter we set. Our retention claim rests instead on an accidental approximation of that controlled comparison. The high-retention methods preserve the answer at near-identical rates (95--99\%) while differing sharply in representational dimensions: rewording, prose-to-list restructuring, token-level pruning, and compression to 2\% of the original length. Within this group retention is approximately constant, so it cannot account for any accuracy differences among these methods; that accuracy nonetheless stays near baseline despite large differences in representational dimensions indicates they have little effect. This isolates the effect of representational dimensions, but only in the high-retention regime. Where retention is lower, representational dimensions and retention vary together, so an accuracy drop cannot be attributed to one rather than the other, and their role there is unmeasured. We can therefore claim only that \emph{among high-retention methods} representational dimensions have limited effect on accuracy; establishing their role across the full range of retention would require constructing representations matched on retention but differing in representational dimensions, and varying them independently.

\subsection{Representational dimensions are not decomposed}
\label{sec:lim-decomp}
The dimensions of representation co-vary across our methods --- paraphrase alters wording, propositions alters structure and length, LLMLingua2 alters coherence and length --- so we vary these dimensions as a bundle rather than isolating them individually, and cannot determine whether they act independently or interact. Our results therefore speak to whether particular, recognisable representations preserve accuracy at high retention, not to which individual dimensions are responsible. Establishing the latter would require representations that vary one dimension while holding the others, and retention, fixed.

\subsection{Scope}
\label{sec:lim-scope}
Our evidence comes from a single dataset (KILT-NQ), a single task type (single-hop, short-answer question answering), and four open generators in the 8--12B range. Each bounds how far the retention finding should be taken. The dataset choice has a specific consequence for one method: RECOMP-extractive-50's relevance scorer is a bi-encoder trained on Natural Questions, the source of KILT-NQ, so its high retention (96.9\%) may partly reflect in-domain advantage that would not transfer to other query distributions.

Every representation we evaluate is natural-language text: the retrieved document is transformed into another readable document, and both our accuracy and retention measures presuppose this, since the LLM judge reads the transformed text to determine whether the answer survives. A separate line of work represents retrieved content not as text but as latent vectors or compressed key-value states consumed directly by the generator \citep{cheng_xrag_2024, rau_context_2025}. These are a more radical departure in representation than any transformation we test, yet our framework cannot assess them: when the document never exists as readable text, per-document answer retention is not measurable in the same terms. Our finding that representation has limited effect when retention is high therefore concerns \emph{textual} representations specifically. Whether it extends to latent representations is an open question, and one our method is not equipped to answer.

Within textual representations, the nine transformation methods we evaluate are exemplars of established techniques rather than a designed sweep of the space: within each category we sample particular methods rather than spanning the range of forms each could take. A textual representation unlike those we test could behave differently --- we do not claim to have characterised the full space of textual representations.

The task type is the most consequential. Retention is well-defined as a per-document property precisely because the task is single-hop: the answer is contained within a single gold document, so asking whether that document still supports it after transformation is a clean question. Under multi-hop reasoning the answer must be composed across several documents, and per-document retention ceases to be well-defined; representation-level properties such as the cross-passage logical connectivity that \citet{chang_what_2025} identify may then carry weight they do not here. Our finding may therefore hold most clearly in the regime least demanding of cross-document structure, and we do not claim it transfers to multi-hop or long-form generation.

Model scale is a related limit. All four generators are small open models; parametric knowledge and context-handling both vary with scale, and we cannot say whether the dominance of retention persists for substantially larger generators.

\subsection{Idealised retrieval conditions}
\label{sec:lim-goldinjection}

Gold injection guarantees an answer-bearing document in every retrieved set (\S\ref{subsec:dataset}); together with per-answer provenance, this makes per-document retention well-defined. But it also means our study characterises the representation stage \emph{conditional on retrieval having surfaced the answer}. The relative importance of representation versus retrieval quality is out of scope by construction. In deployment, where retrieval often fails to surface an answer-bearing document at all, the balance of factors will differ from what we report.

\subsection{Limits of the retention account}
\label{sec:subsec-d4}
Retention is the strongest predictor of generator accuracy in our results, but not the only factor. Three qualifications limit the retention account: (1) accuracy losses are smaller than retention losses, (2) Mistral-Nemo's behaviour is not well predicted by retention, and (3) both retention and accuracy are measured through an LLM judge whose sensitivities overlap with those of the generators it evaluates.

The first qualification is the accuracy--retention gap in Figure~\ref{fig:accuracy_v_retention}: for methods with substantial retention loss, accuracy losses are markedly smaller than retention losses. Parametric knowledge and answer content in non-gold retrieved documents give the generator routes to a correct answer that gold-document retention does not capture. Whether these fully account for the gap is unclear.

The second qualification concerns Mistral-Nemo, whose behaviour falls outside the retention account. Its closed-book accuracy and gold-only accuracy fall within the range of the other generators, indicating that its parametric knowledge and single-document handling are intact. The deficit is specific to RAG: accuracy on the original retrieved set is 73.9\%, 6--7 points below Qwen, Gemma, and Llama. Two comparisons isolate components of Mistral's deficit. Gold-5x, which holds information content fixed while adding length and repetition, costs Mistral 4 points relative to gold-only (80.8\% vs 84.8\%), reflecting sensitivity to length or repetition (our design does not separate the two). Replacing four of those copies with retrieved documents costs a further 7 points (73.9\% vs 80.8\%), reflecting the cost of substituting non-gold content. This points to a length or repetition sensitivity, but the transformations do not corroborate it: size does not track accuracy in either direction. The methods that significantly exceed baseline span almost the entire size range (3\% to 103\%), while the two size extremes --- propositions at 120\% and snippet-abstractive-Llama at 2\% --- are both statistically indistinguishable from it. The property driving Mistral's deficit is therefore unidentified: it is neither answer retention nor length alone, and it surfaces only when the full retrieved set is present. Generalising the retention account will require isolating that context-level factor and establishing which generators are sensitive to it.

A third qualification concerns the measurement instrument itself. Both retention and answer accuracy are assessed by the same language model, Qwen 2.5 32B. Because the judge is itself a language model doing the same kind of answer-extraction as the generators it evaluates, it inherits the same sensitivities to input representation. A reformulation that repositions an answer-bearing fact, or restructures it from prose into a list, may be judged to have lower retention or lower accuracy not because the answer is harder for the generator to use, but because it is harder for the judge to detect.

%% file: appendix.tex
\section{Prompts}
\label{app:prompts}

\subsection{Transformation}
\label{app:prompts-transform}

\newtcolorbox{promptbox}[1][]{
	breakable, enhanced, sharp corners,
	colback=gray!5, colframe=gray!50, boxrule=0.5pt,
	fonttitle=\bfseries\small, coltitle=black,
	fontupper=\small\ttfamily,
	fontlower=\small\ttfamily,
	left=3pt, right=3pt, top=3pt, bottom=3pt, #1
}

% summary
\begin{figure}[H]
	\begin{promptbox}[title=Section $\Rightarrow$ Summary]
		You are summarizing a section from a Wikipedia article.
		
		Task: Summarize the provided section while preserving key facts.\\
		- Use ONLY information stated in the section.\\
		- Preserve named entities, dates, and numbers exactly as written.\\
		- Do not infer, generalize, or add new facts.\\[4pt]
		Output ONLY the summary.
		
		\tcblower
		Section: \{section\}
		
		Summary:
	\end{promptbox}
	\caption{Summary transformation prompt. \texttt{\{section\}} is filled per section.}
	\label{fig:prompt-summary}
\end{figure}

% snippet-abstractive
\begin{figure}[H]
	\begin{promptbox}[title=Document + Query $\Rightarrow$ Abstract]
		You are creating a query-focused abstract of a document.
		
		Task: In 100 words or fewer, write a concise answer-style abstract that helps answer the question using ONLY this document.\\
		- Include ONLY information explicitly stated in the document. No outside knowledge.\\
		- You MAY paraphrase for clarity, but every claim must be directly supported by the document.\\
		- Include only useful points for answering the question; omit everything else.\\
		- Preserve named entities, dates, and numbers exactly as written.\\
		- If the document contains no relevant information, write ``No relevant information found.''\\[4pt]
		Output ONLY the abstract. No preamble, apologies, or commentary.
		
		\tcblower
		Question: \{query\}
		
		Document: \{retrieved\_doc\}
		
		Query-focused abstract:
	\end{promptbox}
	\caption{Snippet-abstractive transformation prompt. \texttt{\{query\}} and \texttt{\{retrieved\_doc\}} are filled per query.}
	\label{fig:prompt-snippet-abstractive}
\end{figure}
%\end{figure}

% paraphrase
\begin{figure}[H]
	\begin{promptbox}[title=Paragraph $\Rightarrow$ Paraphrase]
		Paraphrase the provided text, rephrasing the content in your own words while preserving all information. Include only information stated in the original text; do not add new facts or interpretations. Write only in English.
		
		\tcblower
		Text: \{retrieved\_doc\}
		
		Paraphrase:
	\end{promptbox}
	\caption{Paraphrase transformation prompt. \texttt{\{retrieved\_doc\}} is filled per document.}
	\label{fig:prompt-paraphrase}
\end{figure}

% proposition
\begin{figure}[H]
	\begin{promptbox}[title=Section $\Rightarrow$ Propositions]
		Decompose the ``Content'' into clear and simple propositions, ensuring they are interpretable out of context.\\
		1. Split compound sentence into simple sentences. Maintain the original phrasing from the input whenever possible.\\
		2. For any named entity that is accompanied by additional descriptive information, separate this information into its own distinct proposition.\\
		3. Decontextualize the proposition by adding necessary modifier to nouns or entire sentences and replacing pronouns (e.g., ``it'', ``he'', ``she'', ``they'', ``this'', ``that'') with the full name of the entities they refer to.\\
		4. Present the results as a bulleted list. Each line must start with ``- ''.\\[4pt]
		Output ONLY the propositions. Do not include any preamble, commentary, or bullet styles other than ``- ''.\\[6pt]
		Example:\\[4pt]
		Title: \={E}ostre\\
		Section: Theories and interpretations, Connection to Easter Hares\\[4pt]
		Content:\\
		The earliest evidence for the Easter Hare (Osterhase) was recorded in south-west Germany in 1678 by the professor of medicine Georg Franck von Franckenau, but it remained unknown in other parts of Germany until the 18th century. Scholar Richard Sermon writes that ``hares were frequently seen in gardens in spring, and thus may have served as a convenient explanation for the origin of the colored eggs hidden there for children. Alternatively, there is a European tradition that hares laid eggs, since a hare's scratch or form and a lapwing's nest look very similar, and both occur on grassland and are first seen in the spring. In the nineteenth century the influence of Easter cards, toys, and books was to make the Easter Hare/Rabbit popular throughout Europe. German immigrants then exported the custom to Britain and America where it evolved into the Easter Bunny.''\\[4pt]
		Propositions:\\
		- The earliest evidence for the Easter Hare was recorded in south-west Germany in 1678 by Georg Franck von Franckenau.\\
		- Georg Franck von Franckenau was a professor of medicine.\\
		- The evidence for the Easter Hare remained unknown in other parts of Germany until the 18th century.\\
		- Richard Sermon was a scholar.\\
		- Richard Sermon writes a hypothesis about the possible explanation for the connection between hares and the tradition during Easter.\\
		- Hares were frequently seen in gardens in spring.\\
		- Hares may have served as a convenient explanation for the origin of the colored eggs hidden in gardens for children.\\
		- There is a European tradition that hares laid eggs.\\
		- A hare's scratch or form and a lapwing's nest look very similar.\\
		- Both hares and lapwing's nests occur on grassland and are first seen in the spring.\\
		- In the nineteenth century the influence of Easter cards, toys, and books was to make the Easter Hare/Rabbit popular throughout Europe.\\
		- German immigrants exported the custom of the Easter Hare/Rabbit to Britain and America.\\
		- The custom of the Easter Hare/Rabbit evolved into the Easter Bunny in Britain and America.
		
		\tcblower
		Title: \{title\}\\
		Section: \{section\_header\}
		
		Content:\\
		\{section\}
		
		Propositions:
	\end{promptbox}
	\caption{Propositions transformation prompt, adapted from the proposition decomposition prompt of \citet{chen_dense_2024} (Figure~8). We change the output format from a JSON array to a bulleted list and place title, section, and content on separate lines; the one-shot example is reproduced from their prompt. \texttt{\{title\}}, \texttt{\{section\_header\}}, and \texttt{\{section\}} are filled per section.}
	\label{fig:prompt-propositions}
\end{figure}

% generation prompts
\subsection{Generation}
\label{app:prompts-generator}

	% closed book
	\begin{figure}[H]
		\begin{promptbox}[title=Question $\Rightarrow$ Answer (closed-book)]
			Answer the question. Give the shortest correct answer possible. Output only the answer.
			
			\tcblower
			Question: \{question\}
			
			Answer:
		\end{promptbox}
		\caption{Closed-book generator prompt. \texttt{\{question\}} is filled per query.}
		\label{fig:prompt-closedbook}
	\end{figure}

	% RAG
	\begin{figure}[H]
		\begin{promptbox}[title=Documents + Question $\Rightarrow$ Answer (RAG)]
			Answer the question based on the provided documents. Give the shortest correct answer possible. Output only the answer.
			
			\tcblower
			Documents:\\
			\{context\}
			
			Question: \{query\}
			
			Answer:
		\end{promptbox}
		\caption{RAG generator prompt. \texttt{\{context\}} is the concatenated document representations in retrieval-rank order, separated by \texttt{\textbackslash n\textbackslash n-{}-{}-\textbackslash n\textbackslash n}; \texttt{\{query\}} is the question.}
		\label{fig:prompt-rag}
	\end{figure}

% llm judge prompts
\subsection{Judge}
\label{app:prompts-judge}

% answer accuracy
\begin{figure}[H]
	\begin{promptbox}[title=Generated + Gold $\Rightarrow$ Correct / Incorrect]
		You are an expert evaluator for question-answering systems. Your task is to compare a generated answer to the gold (correct) answer(s) and determine if it is correct.\\[4pt]
		A generated answer is CORRECT if ANY of the following apply:\\
		- It conveys the same core information as any of the gold answers\\
		- It is more specific than a gold answer (e.g., ``Barack Obama'' when gold is ``Obama'')\\
		- It is a valid alternative form of a gold answer (e.g., ``USA'' vs ``United States'')\\
		- It includes a gold answer along with additional information\\
		- It lists multiple items where one of them matches a gold answer\\
		- It provides a more complete answer than a gold answer\\[4pt]
		A generated answer is INCORRECT if ANY of the following apply:\\
		- It does not contain any gold answer or its equivalent anywhere in the response\\
		- It contradicts all gold answers\\
		- It is a NO-RESPONSE or empty answer\\[4pt]
		When in doubt, if any gold answer appears within the generated answer, respond CORRECT.\\[4pt]
		Respond with ONLY one word: CORRECT or INCORRECT.
		
		\tcblower
		Question: \{query\}
		
		Generated Answer: \{generated\_answer\}
		
		Gold Answer(s):\\
		\{gold\_answer\}
		
		Evaluation (CORRECT or INCORRECT):
	\end{promptbox}
	\caption{LLM-judge accuracy evaluation prompt. \texttt{\{gold\_answer\}} is the set of accepted answers, each on its own line as a bulleted list; \texttt{\{query\}} and \texttt{\{generated\_answer\}} are filled per evaluation.}
	\label{fig:prompt-judge-accuracy}
\end{figure}

% answer retention
\begin{figure}[H]
	\begin{promptbox}[title=Answer + Document $\Rightarrow$ Present / Absent]
		You are an expert evaluator. Your task is to determine whether any of the accepted answers appear in a document, either exactly or in an equivalent form (e.g.\ partial names, alternate date formats, abbreviations).\\[4pt]
		PRESENT: At least one accepted answer appears in the document.\\
		ABSENT: None of the accepted answers appear in the document.\\[4pt]
		Respond with ONLY one word: PRESENT or ABSENT.
		
		\tcblower
		Question: \{question\}
		
		Accepted answers:\\
		\{gold\_answer\}
		
		Document:\\
		\{document\}
		
		Is the answer present in the document? (PRESENT or ABSENT):
	\end{promptbox}
\caption{LLM-judge answer-retention prompt, applied to the transformed gold document. \texttt{\{gold\_answer\}} is the set of accepted answers, each on its own line as a bulleted list; \texttt{\{question\}} and \texttt{\{document\}} are filled per evaluation.}
	\label{fig:prompt-judge-retention}
\end{figure}